\documentclass[12pt]{article}
\usepackage{amsmath}
\usepackage{amssymb}
\usepackage{graphicx}
\usepackage{bm}
\usepackage{upgreek}

\begin{document}

{\noindent \bf \Large Explanation for the transverse radiation force observed on a vertically hanging fiber}
\vspace{1cm}
\begin{center}
Iver Brevik\footnote{iver.h.brevik@ntnu.no}

\bigskip
Department of Energy and Process Engineering,\\
Norwegian University of Science and Technology, Trondheim, Norway

\bigskip

\end{center}

\begin{abstract}

As shown in the experiment of She {\it et al.} [Phys. Rev. Lett. {\bf 101}, 243601 (2008)], a weak laser beam sent through a vertically hanging fiber exerts a transverse force and produces a lateral displacement of the fiber's lower end. The experiment is of obvious  theoretical interest in connection with the electromagnetic theory of media.  Suggested explanations given for this effect in the past include the famous Abraham-Minkowski issue concerning the "correct"   photon momentum in matter. In our opinion such an explanation can hardly be right. Instead, we propose  instead a very simple  description of the effect implying that the sideways deflection is caused by the radiation force  on the {\it obliquely cut } lower end face of the fiber.  From a  calculation based upon geometrical optics, we find quite good agreement with the observations. We present also, as an alternative approach, a  calculation involving wave optics instead of geometrical optics, and find comparable results.
\end{abstract}

\section{Introduction}

Consider a straight vertically hanging silica glass fiber of length $L=1.5~$mm through which  low-intensity laser light of power $P=6.4$ mW is sent from above. This was the basic setup of the experiment of She {\it et al.} \cite{she08}. The striking outcome of the experiment was that a {\it sideways} deflection of the lower end, of magnitude $\Delta x=10~\mu$m, was observed.

This observation presents a challenge to conventional electromagnetic theory of dielectric media: how can such an effect be explained? One might think about several options:

First, and this was actually the explanation favored by the experimentalists themselves, one might think that the effect has a bearing on the one-hundred old Abraham-Minkowski energy-momentum problem in dielectrics. In our opinion this explanation is hardly correct, for the following reason: The surface force on the fiber is caused by integration across the surface of the Abraham-Minkowski volume force density ${\bf f}^{\rm AM}=-(\varepsilon_0/2)E^2{\bf \nabla}n^2$ in the surface layer, $n$ being the refractive index. This expression is not related to electromagnetic momentum. One may here recall that the full electromagnetic force density (ignoring electrostriction) is given as
\begin{equation}
{\bf f}= {\bf f}^{\rm AM}+\frac{n^2-1}{c^2}\frac{\partial}{\partial t}\rm \bf (E\times H), \label{1}
\end{equation}
and electromagnetic momentum does not occur until the second term (called the Abraham term) in this expression. See also the comments in Ref.~\cite{brevik09} on this point.

A second possibility is that the sideways motion is caused by impurities in the material, introduced in the mechanical drawing process, making the fiber non- axisymmetric. We do not go further into this point, as it is not easily described quantitatively.

Third, one might envisage that the effect is due to a uniform imbalance $\Delta n$ of the refractive index in the fiber. This idea, which was discussed in Ref.~\cite{brevik10}, is unfortunately incorrect, most simply understood from symmetry arguments; cf. also Ref.~\cite{torchigin13}.

Finally - and this is our main topic below - we shall consider in some detail a most natural explanation of the effect, namely that it is caused by the transverse radiation force that acts when the lower end face of the fiber is not cut exactly at $90^\circ$. It will turn out, even in a simplified geometrical optics approach, that the theory fits very well with the observations.

\section{Transverse optical force: Geometrical optics}

Assume the vertically hanging glass fiber whose lower end has been obliquely cut so that it makes an angle of inclination $\theta$ with the horizontal plane. The refractive index of the fiber is assumed constant, equal to  $n=\sqrt \varepsilon$ (we write the constitutive relations in the form ${\bf D}=\varepsilon_0 \varepsilon \bf E $, ${\bf B}=\mu_0 \bf H$, so that $\varepsilon$ is a nondimensional quantity). A cw laser beam of total power $P$ is sent vertically through the fiber. On the outside there is air (vacuum). We   assume that the fiber has  a circular cross section with radius $R$, and as a first approximation we  apply in this section  geometric optics to the fields in the interior. It is evident  that $\theta$ must be the same as the angle of incidence of the incident wave vector ${\bf k}$. Let the angle of transmission be $\theta_t$. The plane of incidence is formed by the vectors ${\bf k}$ and the outward normal $\bf n$. The incident plane wave is assumed monochromatic, ${\bf E}^{(i)}={\bf E}_0({\bf r})e^{-i\omega t}$, where without loss of generality  ${\bf E}_0(\bf r)$ can be taken to be real. It is convenient to work with the  rms-value of the real electric field; calling this  $\bf E(r)$, we have ${\bf E(r)}=(1/\sqrt 2){\bf E}_0(\bf r)$. For the energy density $w$ and the intensity $S$ of the incident field it then follows that
\begin{equation}
w=\varepsilon_0n^2 E^2, \quad S=\varepsilon_0 ncE^2. \label{2}
\end{equation}
Let ${\bf E}_\parallel$ and ${\bf E}_\perp$ be the components of $\bf E$ parallel and perpendicular to the plane of incidence,
\begin{equation}
 {\bf E}_\parallel =E\cos \alpha, \quad {\bf E}_\perp =E\sin \alpha. \label{23}
 \end{equation}
 The respective transmission coefficients are
 \begin{equation}
 T_\parallel =\frac{ \sin 2\theta\sin 2\theta_t}{\sin^2(\theta+\theta_t)\cos^2(\theta-\theta_t)}, \label{4}
 \end{equation}
 \begin{equation}
 T_\perp =\frac{\sin 2\theta\sin 2\theta_t}{\sin^2(\theta+\theta_t)}. \label{5}
 \end{equation}
 The surface force density (radiation pressure) at the lower end of the fiber is found by integrating the Abraham-Minkowski force density ${\bf f}^{\rm AM}=-\frac{1}{2}\varepsilon_0 E^2{\bf \nabla}\varepsilon $ in the  boundary region across the boundary. Calling this pressure  ${\bf \sigma}^{\rm AM}$, we have
 \begin{equation}
 {\bf \sigma}^{\rm AM}=\frac{S(n^2-1)}{2c}\frac{\cos \theta}{\cos \theta_t}\left[ (\sin^2\theta+\cos^2 \theta_t)T_\parallel\cos ^2\alpha+T_\perp\sin^2 \alpha\right] \bf n \label{6}
 \end{equation}
 (derivations of this result can be found, for instance, in Refs.~\cite{aanensen13}, \cite{delville06} or \cite{hallanger05}).

 This expression has the virtue that it is general, valid for all values of $\theta$ and $\alpha$. When applied to the experiment of She {\it et al}, we expect that the angle $\theta$ is small, $\theta \ll 1$.  Moreover, for simplifying reasons we take the incident  wave to be polarized in the plane of incidence so that $\alpha=0$. Then the formula (\ref{6}) becomes significantly reduced,
 \begin{equation}
 {\bf \sigma}^{\rm AM}= \frac{S(n^2-1)}{2c}T_\parallel \,{\bf n}=\frac{2nS}{c}\frac{n-1}{n+1}\bf n. \label{7}
 \end{equation}
 Multiplying this with the cross-sectional area $\pi R^2$ and taking the component transverse to the fiber we obtain the bending force, hereafter called $F_T$, as
 \begin{equation}
 F_T=\frac{2nP}{c} \frac{n-1}{n+1}\sin \theta, \label{8}
 \end{equation}
 where $P=S\pi R^2$ is the incident power.

 Let us next apply elasticity theory to the fiber,  modeling it as a rod with  Young's modulus equal to $E$ and an area moment of inertia about the centroidal axis equal to $I$. The rod  is clamped at $z=0$ and subject to the transverse bending force   $F_T$ in the $x$ direction at the tip $z=L$. According to the observations in the experiment of She {\it et al.}, the bending $\Delta x$ at the tip is small in comparison with $L$ (about $10~\mu$m compared to 1.5 mm). This means that we can make use of the governing equation for {\it weak} bendings, $z'''=-F_T/EI$, which is to be integrated  subject to the boundary conditions $x=0, x'=0$ at $z=0$ and $z''=0$ at $z=L$. The analytic form of the rod then becomes  \cite{landau70}
\begin{equation}
x(z)=\frac{F_T z^2}{6EI}(3L-z), \quad x(L) \equiv \Delta x= \frac{F_TL^3}{3EI}. \label{9}
\end{equation}
Eliminating $F_T$ from Eqs.~(\ref{8}) and (\ref{9}) we can now express the angle of inclination $\theta_i$ in terms of measurable quantities,
\begin{equation}
\sin \theta=\frac{3EIc}{2L^2 P}\,\frac{n+1}{n(n-1)}\,\frac{\Delta x}{L}. \label{10}
\end{equation}
For glass, it is known that the Young's modulus lies between 50 and 90 GPa. Let us choose $E=70$ GPa, as in \cite{she08}. For a cylindrical cross section, the moment of inertia is
\begin{equation}
I=\frac{\pi}{4}R^4. \label{11}
\end{equation}
Further, following \cite{she08} we take the fiber width to be $2R=0.45~\mu$m, and insert $L=1.5~$mm, $P=6.4~$mW, $n=1.46$. Equation (\ref{10}) then yields, in order to correspond to a lateral displacement of $\Delta x=9~\mu$m,
\begin{equation}
\sin \theta=0.0971, \quad \theta \approx 6^\circ. \label{12}
\end{equation}
The agreement with the observations is better than one might expect; it was  estimated in \cite{she08} that the end face of the fiber had an angle of inclination of about $8^\circ$.  As the value of $\theta$ is so small, our simplification above in going from Eq.~(\ref{6}) to Eq.~(\ref{7}) is justified.

\section{Use of wave theory}

Although or results above are supportive, one should bear in mind that the calculation is based upon geometrical optics. It might be that this approximation is somewhat crude, as the transverse dimension of the fiber in the experiment was small  (this point has been emphasized also by  Mansuripur  \cite{mansuripur09}). It is  natural therefore, as an alternative,  to approach the problem from a different angle implying  use of  wave optics instead. Let us in the following consider a simplified planar model of the fiber where it is taken to be a uniform slab, in the ideal limit infinite in the horizontal $y$ direction, having a finite width $2a$ in the other horizontal $x$ direction. The laser light is sent downward in the vertical $z$ direction such as above. The symmetry plane is at $x=0$ (for a figure sketch, cf. Fig. 2.1 in Ref.~\cite{okamoto00}, which is essentially reproduced as  Fig.~1 in Ref.~\cite{brevik10}). Outside the fiber, for $x>a$ and $x<-a$, we assume a vacuum (air).

Assume that the incident  electric field is polarized in the $y$ direction, ${\bf E}^{(i)}=(0,E_y,0)$. One has in this case, when omitting the common factor $\cos(\beta z-\omega t)$ with $\beta$ the axial wave number \cite{okamoto00,brevik10},
\begin{eqnarray}
E_y=\left\{ \begin{array}{lll}
A\cos \kappa a \,e^{-\xi (x-a)}, & x>a, \\
A\cos \kappa x, & -a \leq x \leq a, \\
A\cos \kappa a \,e^{\xi (x+a)}, & x<-a.
\end{array}
\right.  \label{13}
\end{eqnarray}
Here
\begin{equation}
\kappa=\sqrt{(n^2\omega^2/c^2)-\beta^2}, \quad \xi=\sqrt{\beta^2-(\omega^2/c^2)}, \label{14}
\end{equation}
and the nondimensional transverse wave vectors are $u=\kappa a$ and $w=\xi a$. These satisfy the relationships
\begin{equation}
u=\arctan \left(\frac{w}{u}\right), \quad u^2+w^2=\frac{\omega^2a^2}{c^2}(n^2-1), \label{15}
\end{equation}
permitting us to determine the values of $\kappa, \xi$, and hence the wave number $\beta$.

The expressions (\ref{13}) refer to the lowest order mode where the phase angle, conventionally called $\phi$, is set equal to zero
 (cf. Ref.~\cite{okamoto00}).

 The quantity $A$ in Eq.~(\ref{13}) is related to the total power  $P$  transmitted by the fiber. If $b$ designates the fiber width in the $y$ direction, one has (cf. formula (2.34) in Ref.~\cite{okamoto00})
\begin{equation}
A^2=\frac{2\omega \mu_0P}{\beta ab[1+(1/w)]}. \label{16}
\end{equation}
It is convenient to define $n_e$ as the 'refractive index' corresponding to the axial wave number $\beta$,
\begin{equation}
\beta =n_e \omega/c. \label{17}
\end{equation}
The cross-sectional area of the model fiber is $2ab$. We shall assume  $b=2a$ in the following (admittedly a rough approximation in a planar waveguide theory), so that the cross section becomes quadratic with area $4a^2$.

What to choose for the linear dimension $2a$ is also subject to some choice. One might choose $2a=2R=0.45~\mu$m, i.e., the same as the diameter in the preceding circular case. Here, will however  determine the value of $a$ from requiring that  the  cross-sectional area should  the same in the two cases. From the condition $4a^2=\pi R^2$ we obtain $2a=0.40~\mu$m, i.e.,  a slightly smaller value.

Further  inserting $\lambda=0.65~\mu$m, $n=1.46$ we now find for the lowest order mode
\begin{equation}
n_e=1.372.  \label{18}
\end{equation}
From Eq.~(\ref{15}) we calculate the nondimensional transverse wave numbers $u=0.966, w=1.816$, and inserting
 $P=6.4$ mW as before  we find from Eqs.~(\ref{16}) and (\ref{17})
\begin{equation}
A=\frac{1}{a}\left[\frac{\mu_0cP}{n_e(1+1/w)}\right]^{1/2}= 8.25\times 10^6~\rm \frac{V}{m}.
\label{19}
\end{equation}
We now calculate the normal surface force density $\sigma^{\rm AM}$ at the end of the fiber in the same way as above, by multiplying $S(n^2-1)/2c$ with the transmission coefficient $T_\parallel =4n/(n+1)^2$. The transverse force $F_T$ thereafter follows by multiplying with $\sin \theta$. This is a valid procedure as long as $\theta$ is small. Thus Eq.~(\ref{7}) still holds. What is needed is to calculate  the magnitude $S$ of Poynting's vector  and integrate it over the fiber cross section.  Within the fiber we have the complete expressions $E_y=A\cos \kappa x \cos(\beta z-\omega t), H_x=-(\beta/\mu_0\omega)E_y$, from which we calculate, after averaging over an optical  period,
\begin{equation}
S= \varepsilon_0n_ecE_y^2=\frac{1}{2}\varepsilon_0n_ecA^2\cos^2\kappa x. \label{20}
\end{equation}
The  integral of $S$ over the cross section thus becomes
\begin{equation}
2a\int_{-a}^a Sdx= \frac{1}{2}\varepsilon_0n_eca^2A^2\left[ \frac{\sin 2u}{u}+1\right], \label{21}
\end{equation}
(recall that $u=\kappa a$).

Starting from Eq.~(\ref{7}) we can now write the transverse force as
\begin{equation}
F_T=\varepsilon_0n_ea^2A^2\,\frac{n(n-1)}{n+1}\left[ \frac{\sin 2u}{u}+1\right]\sin \theta. \label{22}
\end{equation}
Here relating $A^2$ to the power $P$ via Eq.~(\ref{16}), and then equating (\ref{22}) to the elasticity-theory expression for $F_T$ found from Eq.~(\ref{9}), we obtain
\begin{equation}
\sin \theta = \frac{3EIc}{L^2P}\,\frac{n+1}{n(n-1)}\,\frac{1+1/w}{(\sin 2u)/u+1}\,\frac{\Delta x}{L}. \label{23}
\end{equation}
We see that this expression is not very different from the expression (\ref{10}) found above. The wave-theory expression contains the transverse nondimensional wave numbers $u$ and $w$, as expected. Moreover, as we are now dealing with a rectangular rod of base $b=2a$ and height $2a$, we have to replace the former expression (\ref{11}) for the moment of inertia $I$ with
\begin{equation}
I=\frac{1}{12}b(2a)^3= \frac{4}{3}a^4. \label{24}
\end{equation}
 Recalling for convenience the input numbers, $L=1.5$ mm, $ P=6.4$ mW, $n=1.46$, $2a=0.40~\mu$m, $E$=70 GPa, $u=0.966, w=1.816$,  we obtain from Eqs.~(\ref{23}) and (\ref{24}), requiring $\Delta x$ to be $9~\mu$m as before,
 \begin{equation}
 \sin \theta=0.161, \quad \theta \approx 9^\circ. \label{25}
 \end{equation}
 Again, it is seen that the agreement with the estimates from the experiment is very satisfactory. A word of warning is however needed here, as the wave-theory approach based upon a rectangular cross section is very sensitive for the choice made for the width. Adopting the value $2a=0.45~\mu$m (actually the first option discussed after Eq.~(\ref{17})) we would instead calculate the numbers $n_e=1.353, u=1.084, w=2.044$, resulting in the final result $\sin \theta=0.278$ $(\theta \approx 16^\circ)$. This value is most likely too high, and it would even violate the approximations leading to Eq.~(\ref{7}).  The main reason for this difference lies in the dependence of the moment of inertia with respect to $a$, $I \propto a^4$. For this reason we suggest that the circular-geometry approach in Sec.~2 is after all the safest one, in spite of its limitation to  geometrical optics.

 \section{Conclusion}

 The sideways motion of a vertically hanging fiber transmitted by a laser beam, as observed by She {\it et al.} \cite{she08} can be explained in a simple and natural way as a consequence of an obliquely cut fiber end. The angle of inclination $\theta$ needs only to be small, probably less than $10^\circ$, in order to correspond to a transverse deflection $\Delta x$ of about $10~\mu$m. We followed  two different approaches, one dealing with geometrical optics, the other one dealing with optical wave theory, and the  results were comparable and in agreement with observations.

 Finally, one may ask if this mechanical action of optical forces can be of practical use. A straightforward application of the effect considered above seems hardly feasible,  but the general possibility  of manipulating micron-scale devices by means of optical laser forces definitely needs attention.

\end{document}